\documentclass[tradiabstract]{aa}
\usepackage{natbib}
\usepackage{txfonts}
\usepackage{graphicx} 
\bibpunct{(}{)}{;}{a}{}{,}
\usepackage[english]{babel}
\usepackage{ulem}

\newcommand{\ms}{\mbox {$\rm M_{\odot}$}}
\newcommand{\ace}{\mbox {$\alpha_{\rm CE}$}}
\newcommand{\md}{\mbox {$\dot{M}$}}

\newcommand{\myr}{\mbox {~${\rm M_{\odot}~yr^{-1}}$}}

\newcommand{\al}{\mbox {$\alpha_{\rm CE} \times \lambda$}}
\newcommand{\te}{\mbox {$T_{\rm eff}$}}
\newcommand{\per}{\mbox {$P_{\rm orb}$}}
\newcommand{\mbh}{\mbox {$M_{\rm bh}$}}

\def\apgt{\ {\raise-.5ex\hbox{$\buildrel>\over\sim$}}\ }
\def\aplt{\ {\raise-.5ex\hbox{$\buildrel<\over\sim$}}\ }

\def\spose#1{\hbox to 0pt{#1\hss}}
\def\simless{\mathrel{\spose{\lower 3pt\hbox{$\mathchar"218$}}
        \raise 2.0pt\hbox{$\mathchar"13C$}}}
\def\simgreat{\mathrel{\spose{\lower 3pt\hbox{$\mathchar"218$}}
        \raise 2.0pt\hbox{$\mathchar"13E$}}}
\def\lta{\mathrel{\spose{\lower 3pt\hbox{$\mathchar"218$}}
        \raise 2.0pt\hbox{$\mathchar"13C$}}}
\def\gta{\mathrel{\spose{\lower 3pt\hbox{$\mathchar"218$}}
        \raise 2.0pt\hbox{$\mathchar"13E$}}}

\begin{document}
\title{Evolutionary models of short-period soft X-ray transients: comparison with observations}

\author{L. R. Yungelson\inst{1,2}
        \and
         J.-P. Lasota\inst{1,3}
}


\institute{Institut d'Astrophysique de Paris, UMR 7095 CNRS,
           UPMC Univ Paris 06, 98bis Bd Arago, 75014 Paris,
           France
            \and
        Institute of Astronomy of the Russian Academy of
            Sciences, 48 Pyatniskaya Str., 119017 Moscow, Russia
            \and
            Astronomical Observatory, Jagiellonian University, ul. Orla 171, 30-244
Krak\'ow, Poland
                            }

\date{received \today}

\titlerunning{Evolution of SXT}
\authorrunning{L. R. Yungelson and J.-P. Lasota}

\abstract{ We consider evolutionary models for the population of
short-period (<10 hr) low-mass black-hole binaries
(LMBHBs) and compare them with observations of soft X-ray transients
(SXTs). We show that assuming strongly reduced magnetic braking (as
suggested by us before for low-mass semidetached binaries) the
calculated masses and effective temperatures of secondaries are
encouragingly close to the observed masses and effective
temperatures (as inferred from their spectra) of donor stars in
short-period LMBHBs. Theoretical mass-transfer rates in SXTs are
consistent with the observed ones if one assumes that accretion
discs in these systems are truncated (``leaky''). We find that the
population of short-period SXTs is formed mainly by systems which
had unevolved or slightly evolved main-sequence donors ($M_2 \lesssim
1.2M_\odot$) with a hydrogen abundance in the center $X_c > 0.35$ at
the Roche-lobe overflow (RLOF). Longer period  
(0.5 - 1) day)SXTs might descend from systems with initial donor masses
of about 1 $M_\odot$ and $X_c < 0.35$. Thus, one can explain the
origin of short period LMBHB without invoking donors with cores
almost totally depleted of hydrogen. Our models suggest that, unless
the currently accepted empirical estimates of mass-loss rates by
winds for
 massive O-stars and Wolf-Rayet stars are
significantly over-evaluated, a very high efficiency of
common-envelope ejection is necessary to form
short-period LMBHBs.

}

\maketitle

\section{Introduction}
\label{sec:intro}

\begin{table*}
\flushleft
\caption[]{Estimates of spectral types, effective temperatures of donor-stars
 and mass ratios of components
$M_2/M_{\rm X}$ in SXT.}
\begin{tabular}{llllllll}
\hline\hline
&Object & $P_0,$ & Sp & $T_{\rm eff}$ & $q$ &Ref. & Comments\\
&& hour & & & &\\
\hline
1& XTE J1118+480 (KV UMa) & 4.104 & K7V-M0.5V & & $0.083$ & 1 & \\
& & & K7V & & &2, 3 & \\
& & & K5/K7V & & & 4 & \\
&& & K5-M0 & & &5 & \\
             & & & K7-M0V & & &6 & First IR-observartions\\
& & & mid to late K & $4700\pm 100$ & & 7 & \\
& &&&& $\sim 0.008$ & 8 &\\
&&&K5V-M1V & & $<0.1$ & 9 & \\
&&&&&0.044 - 0.035& 10 &\\
2&GRO J0422+32 (V518 Per) & 5.088 & ${\rm M2\pm2V}$ & & &11 & \\
&&& ${\rm M2^{+2}_{-1}V}$ & & $0.116^{+0.079}_{-0.071}$ &12 & \\
&&& M1V& & &13 & \\
&&& M1V & & &14 & B5 to K7 from $(H-K)_0$\\
&&&&&0.313 - 0.076 & 10 & col. index\\
3&GRS 1009-45 (MM Vel) & 6.840 & later than G5V-K0V & & &15 & \\
&&& K7V-M0V & & &16 &possibly K6V \\
&&&&&0.159 - 0.125 & 10 &\\
4&XTE J1650-500 & 7.680 & K4 V & & 0.1 &17 & next best \\
&&&&&&& match G5V and K2III \\
5&A0620-00 (V616 Mon) & 7.752 & later than a K3V,& & & 18 & \\
&&& most likely&&&&\\
&&& between K5V and K7V & & & &\\
& & & K5 V & & & 19 & \\
& & & K3V & & & 20 & \\
& &&& $4900\pm 100$ && 21 & \\
&&&&&0.075 - 0.055 & 10 &\\
6&GS 2000+25 (QZ Vul) & 8.280 & K5V & & & 22 & evolved \\
& & & K3-K6V & & &23 & slightly evolved,\\
&&&&&&& but not a subgiant\\
& & & consistent with K5V & & &24 & K4V is nearly identical, \\
& & & & & & &G5-K1 and K8-M0 also give \\
&&&&&&&very good correlations\\
&&&&&0.053 - 0.035 & 10 &\\
7&XTE J1859+226 (V406 Vul) & 9.120 & G5-K0 & & &25 & G5V fits best \\
8&GRS 1124-68 (GU Mus) & 10.392 & K5V to K7V & & &26 & \\
& & & K3 -K5V & & &27 & slightly evolved,\\
&&&&&&& K7 features not observed\\
& & & K3-K4V & & &28 & \\
& & & K3/5V & & &29& \\
&&&&&0.208 - 0.114 &10&\\
9& H 1705-25 (V2107 Oph) & 12.504 & K7V & & &30 & K3 to M0 also give good\\
& & & K3V & & &31 & correlations\\
& & & K5V & & &32 & \\
&&&&&$<0.053$ &10&\\
\hline
\end{tabular}
\vskip 0.3cm
References:
1 -- \citet{2001ApJ...556...42W},
2 -- \citet{2002MNRAS.333..791Z},
3 -- \citet{2006ApJ...642..438G},
4 -- \citet{2003ApJ...593..435M},
5 -- \citet{2004ApJ...612.1026T},
6 -- \citet{2005MNRAS.362L..13M},
7 -- \citet{2006ApJ...644L..49G},
8 -- \citet{2001IAUC.7617....1C},
9 -- \citet{2001ApJ...551L.147M},
10 -- \citet{2003IAUS..212..365O},
11 -- \citet{1995MNRAS.276L..35C},
12 -- \citet{harl99},
13 -- \citet{2003ApJ...599.1254G},
14 -- \citet{2007MNRAS.374..657R}
15 -- \citet{della97},
16 -- \citet{1999PASP..111..969F},
17 -- \citet{2004ApJ...616..376O},
18 -- \citet{2007ApJ...663.1215F},
19 -- \citet{2007AJ....133..162H}
20 -- \citet{shah99},
21 -- \citet{2004ApJ...609..988G},
22 -- \citet{2004AJ....127..481I},
23 -- \citet{1996PASP..108..762H},
24 -- \citet{1995ApJ...455L.139F},
25 -- \citet{2001IAUC.7644....2F},
26 -- \citet{1996AJ....111.1675K},
27 -- \citet{1996ApJ...468..380O},
28 -- \citet{1997NewA....1..299C},
29 -- \citet{1997MNRAS.285..607S},
30 -- \citet{1997PASP..109..461F},
31 -- \citet{1996ApJ...459..226R},
32 -- \citet{1997AJ....114.1170H}
\label{tab:pte}
\end{table*}

Soft X-ray transients (SXTs) are a sub-class of low-mass X-ray
binaries most of which harbour black holes \citep[see, e.g.,
][]{2006ARA&A..44...49R}. Their transient behaviour is commonly
associated with the same thermal-viscous instability of accretion
discs that drives outbursts of dwarf-novae \citep[see, e.g.,][ and
references therein]{dubhamlas01,L01}. In this article we will be
interested in black-hole systems with orbital periods shorter than
$\sim0.5$ day. Some observational data on these systems are
summarised in Table \ref{tab:pte}. All known low-mass black-hole
binaries (LMBHBs) are transient.

There are at least two major open questions concerning the origin
and evolution of SXTs. First, the values of the parameters
describing the common envelope phase, second the strength of the
angular momentum loss through magnetic braking.

\subsection{The common-envelope phase}

As first suggested by \citet{macclintock86} and \citet{khp87}
the progenitors of LMBHBs may be relatively wide binaries (but
still ``close'' in an evolutionary sense) composed of a
massive primary ($M_{10} \gta (25-40)$\,\ms) and a low-mass
companion ($M_{20} \lta 1$\,\ms). Such a binary avoids
merging in the common envelope which is
formed when a massive star overflows its critical lobe and
survives the supernova explosion that produces the black hole.
A black hole plus main-sequence star (henceforth, ``bh+ms'')
binary is formed. As in cataclysmic variables, the further
evolution of the system is controlled by the loss of
angular momentum through gravitational radiation and/or
magnetically coupled stellar winds \citep{khp87,pylsav89}. This
evolutionary path for LMBHBs has been challenged on the basis
of computations which showed that envelopes of massive stars
are very tightly bound to their cores.
The ``standard'' equation for the variation of the
orbital separation of components based on the balance between
the binding energy of the mass-losing star and the
orbital energy of the system \citep{web84,khp87} implies a ratio of final $a_f$ to
initial $a_i$ separations of components equal to
\begin{equation}
\label{eq:ce}
\frac{a_f}{a_i}= \frac{M_{1,c}}{M_1}\left[ 1+\left( \frac{2}{\ace\lambda r_{1,L}}\right)
\left( \frac{M_1-M_{1,c}}{M_2}\right) \right]^{-1},
\end{equation}
where $\ace$ is the common envelope ejection efficiency,
$\lambda$\ the parameter of the binding energy of the
stellar envelope, $M_1 $ and $ M_{1,c} $ are initial mass of
mass-losing star and the mass of its remnant, $r_{1,L}$ is
the dimensionless radius of the star at the beginning of
mass transfer, $ M_2 $ is the mass of companion. Formally,
applying Eq.~(\ref{eq:ce})
to the estimate of the outcome of the common envelope stage
one finds that a low-mass secondary is unable to
unbind the envelope of a massive primary
\citep{Podsiadlowski&al03,2006MNRAS.369.1152K}. Thus instead of
forming a short-period binary, the components will merge. As
found by \citet{Podsiadlowski&al03} and
\citet{2006MNRAS.366.1415J}, producing a population of LMBHBs
in a ``standard'' scenario requires the product $\alpha_{\rm
CE} \times \lambda$
to exceed $\sim 0.1$ which they consider as
unrealistic. An alternative ``anomalous magnetic braking
scenario'' \citep{2006MNRAS.366.1415J} suggests that the progenitors of
the donors of SXTs are intermediate-mass ($\gta 2\,\ms$)\
Ap/Bp-stars with anomalously high magnetic field strength. In
this case, after an initial high mass-loss rate stage of
evolution, the secondary of the system turns into a low-mass
star with a long evolutionary lifetime. This scenario
explains the short orbital periods of SXTs. However, in such a
scenario the effective temperatures of the descendants of the
intermediate-mass stars significantly exceed the effective
temperatures of the observed SXT donors. On the other hand
\citet{2006MNRAS.369.1152K} found that a population of
black holes accompanied by low-mass secondaries may be formed
if the rate of winds from WR stars is reduced \textit{ad hoc}.

However, the estimates of both \ace\ and $\lambda$ remain
uncertain. The estimate of \al\ is strongly influenced by
the assumptions about the mass-loss by stellar winds and by the
uncertainty about the role of the internal thermodynamic energy
in unbinding the envelope of the donor
(see for the first suggestion of this source of energy
in the context of formation of planetary nebulae \citet{1967AJ.....72Q.813L}
and
e.g. \citet{1995MNRAS.272..800H,Taudew,soker_harpaz03,Podsiadlowski&al03,2007arXiv0704.0280W}
for further discussion of this issue applied
to common envelopes).

\citet{2003RMxAC..18...24D} carried out 3-D common
envelope modelling for an 1.25\,\ms\ AGB star engulfing 0.1
and 0.2\,\ms\ companions that took into account both
rotation and the interaction between the spiralling-in
component and the donor star. If the ratios $a_f/a_i$
obtained by De Marco et al. are inserted in Eq.~(\ref{eq:ce}),
they correspond to \al\ up to $\simeq 2$. Regretfully, similar
calculations are still absent for massive binaries and
one has to rely on indirect methods for evaluation of \al.

The attempts to find a plausible evolutionary scenario for the
pulsar PSRJ 2145-0750 \citep{1994A&A...291L..39V} and the
results of modelling of the population of binary pulsars (which
also invokes high-mass stars) both favour $\al \simeq 2$
\citep{py98}\footnote{Though one cannot exclude that, because
of the difference in the mass of progenitors of neutron
stars and black holes, \al\ for them may be different.}.
\citet{kal99} in her study of evolutionary parameters of
progenitors of donors in black-hole X-ray binaries found that
the explanation of the origin of the latter systems suggests
$\ace > 1$, implying that sources other than orbital energy
release may be invoked in unbinding common envelopes.
\citet{2002MNRAS.329..897H} in their model for the total
Galactic population of interacting binaries obtained a
subpopulation of transient low-mass black-hole binaries for
$0.5 \le \al \le 1.5$. On the other hand using supernova rates
and empirical estimates of the compact object merger rate
\citet{oshaug08} constrained \al\ to be in the range
0.15-0.5.

The issue of the sources of energy that may increase
\ace\ was discussed by \citet{1993PASP..105.1373I}. Referring
the reader to the original paper, we mention only that, apart
from recombination energy in ionization zones, Iben and
Livio suggested, e.g., dynamo generation of magnetic fields
that may contribute to matter ejection, enhanced nuclear
burning due to injection of fresh matter into nuclear burning
shells by circulation movements that develop in common
envelopes and excitation of non-radial pulsations that may drive
mass loss. None of these mechanism has been
explored as yet.

All estimates of \al\ were obtained under different sets of
assumptions on the evolution of massive stars that are consistent
with our current knowledge of the stellar evolution. Thus it is
still interesting to compare the predictions of various
evolutionary models with observations. In this article we
compare the \citet{yungelson_bh06} model with the observed
properties of secondary stars in SXTs \citep[see
also][]{2008NewAR..51..860Y}.

\subsection{Magnetic braking}
\label{sec:mb}

Yet another problem, noted, e.g., by \citet{kkb96,ef98,mnl99} and
\citet{iv_kal06} is associated with the mechanism of AML by low
donor-mass binaries. One finds that if, following \citet{vz81}, one
assumes that the braking law for \textit{single} field stars $\Omega
\propto t^{-0.5}$ \citep{skumanich72} can be extrapolated over an
order of magnitude in the rotational velocity $v$ (from several 10
to several 100 km/s) to the case of close binary systems and if the
spin-orbit coupling is efficient, the predicted mass-transfer rates
for LMBHBs at orbital periods $\gta 2$\,hr are sufficiently high for
these systems to have stable hot discs. However, such a population
of stable and bright low-mass black-hole X-ray binaries has not been
observed. Also, the \citet{skumanich72} ``law'' is apparently in
conflict with observational data on rotation velocities in young
open clusters \citep{2002ASPC..261...11C,andron03}. According to the
latter study, the time-scale of rotational braking is two orders
of magnitude longer than the one based on the Skumanich law. Also
\citet{vanpar96} noted that the values of \md\ in SXTs are close to
those expected if gravitational wave emission is the sole sink of
angular momentum. Also for cataclysmic variables mass transfer
rates predicted by the Skumanich law based AML disagree with
observations, see, e. g., \citet{ham88} and \citet{ivtaam03}.

In our previous work \citep[][ henceforth, Paper
I]{yungelson_bh06} we found that when the Verbunt \& Zwaan AML
mechanism is allowed to operate after the systems become
semi-detached, one obtains a large number of bright, steady
LMBHBs that clearly are not observed in reality (also P.
Charles, private communication). Therefore we suggested, in
line with observational evidence mentioned above, that in the
semi-detached systems with black-hole accretors, magnetic
braking operates on a much reduced scale (as compared
with the Verbunt \& Zwaan prescription), or that it does not operate at all.
As a test of this hypothesis, we computed a population of
LMBHBs under the assumption that MSW is not operating once the RLOF
occurs and have shown that in this case there remains in the
Galaxy about the same number of such systems as in the case
with active MSW ($\simeq10000$) but all of them are transient,
according to the disc instability model (DIM) criterion of
\citet{dlhc99}.

In the present paper we
extend considerations of the ``no-MSW after RLOF'' model
proposed in Paper I to the case of $\al<2$ and
carry out a detailed comparison of the model with
observations.

\section{The model}
\label{sec:method}

For convenience we remind some basic
information about our calculation of the LMBHB population.

The model of the LMXB population is obtained in two
steps: (i) modeling time-dependent formation of the population
of bh+ms binaries, (ii) tracing the subsequent
evolution of each system. The Galactic ensemble of bh+ms
binaries is computed with the population
synthesis code \textsf{SEBA} \citep{pv96,py98,nyp+01,nyp04}
using 250000 initial binaries with $M_{10} \geq 25$\,\ms. The time-
and position-dependent Galactic star formation history in the
code follows the model of \citet{bp99}; for the inner 3 kpc of
the Galaxy the star formation rate given in the latter study is
doubled to mimic the Galactic bulge (see Figs.~1 and 2 of
\citet{nyp04}). The assumed binarity rate is 50\% (2/3 of the stars in
binaries). The IMF follows \citet{ktg03}, the initial distribution
of semi-major axes of binaries ($a$) is flat in $\log a$
between contact and 10$^6$ R$_\odot$. A flat mass ratio
distribution, and an initial distribution of eccentricities of
orbits $\Xi(e)=2e$ are assumed.

For the common-envelope phase we used Eq.~(\ref{eq:ce}).
We tested the combinations of common envelope ejection
efficiency and stellar envelope binding energy parameters
$\al$ = 2, 0.5, and 0.1 (see below).

Black hole progenitors have $M_{10}=25 - 100$\,\ms. The
relation between MS masses of stars and pre-SN masses generated
by \textsf{SEBA} agrees well with the one obtained by the SSE-code
of \citet{hpt00}, despite differences in the treatment of
stellar winds. The algorithm for the formation of black holes
follows the fall-back scenario \citep{fryer_kal_bh01} with
the assumption of a constant explosion energy of $10^{50}$
ergs, which is within the expected range. Nascent black holes
receive kicks at formation that follow the \citet{pac90} velocity
distribution with $\sigma_v=300$ km~s$^{-1}$, scaled
down with the ratio of the black hole mass to neutron star
mass.

In the next step of modeling, the population of bh+ms binaries
born at different epochs is convolved with the grid of
evolutionary tracks for low-mass components in the binaries
with different combinations of masses of components and
post-circularization (initial) orbital periods (see
Fig.~\ref{fig:grid}). All tracks used in the paper were
computed by an appropriately modified TWIN version (September
2003) of the \citet{egg71} evolutionary code. As mentioned in
\S\ref{sec:mb} the AML via magnetic stellar wind was taken
into account following \citet{vz81}:
\begin{equation}
\dot{J} = -0.5\times 10^{-28}f^{-2}k^{2}\left(\frac{2\pi}{P}\right)^3
M_2 R_2^{4},
\label{eq:msw}
\end{equation}
where $M_2$ is mass of the secondary, $R_2$ -- its radius, $k^2\sim 0.1$ -- its gyration radius,
$P$ -- orbital period,
$f \sim 1$ -- a parameter derived
from observations; it was set to 1.
 For momentum losses
via gravitational wave radiation the standard
\citet{ll71} formula was applied.

The evolution of each system was traced
over a time span from formation to $T=13.5$~Gyr or to the epoch
when the mass ratio of the components of the system became
$q=M_2/M_{\rm bh}=0.02$. At $q \aplt 0.02$ the circularization
radius of the accretion stream becomes larger
than the outer radius of the accretion disc, resonance
phenomena in the disc become important and it remains
unexplored as yet how mass transfer then proceeds. The systems
with $q < 0.02$ have $P_{\rm orb} \apgt 1.5$ hr, mass-transfer
rates $\aplt 10^{-10}$\,\myr\ and it remains to be observed if
mass transfer occurs in them.

For a more detailed description of the input parameters
we refer the reader to Paper I.

\section{The population of progenitors}
\label{sec:prog}

\begin{figure}
\parbox[t]{\columnwidth}
{ \includegraphics[angle=-90,width=0.98\columnwidth]{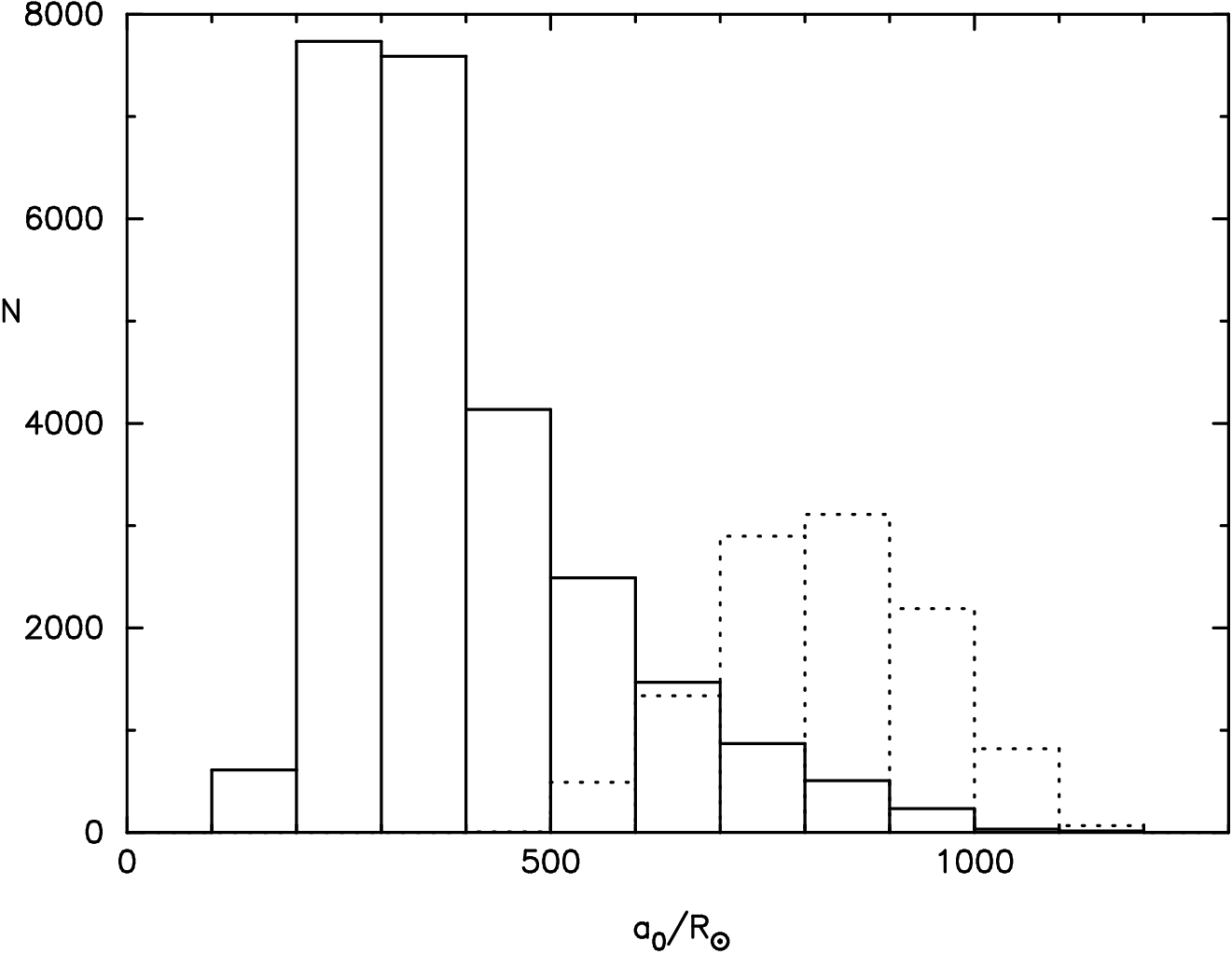}
    \includegraphics[angle=-90,width=0.98\columnwidth]{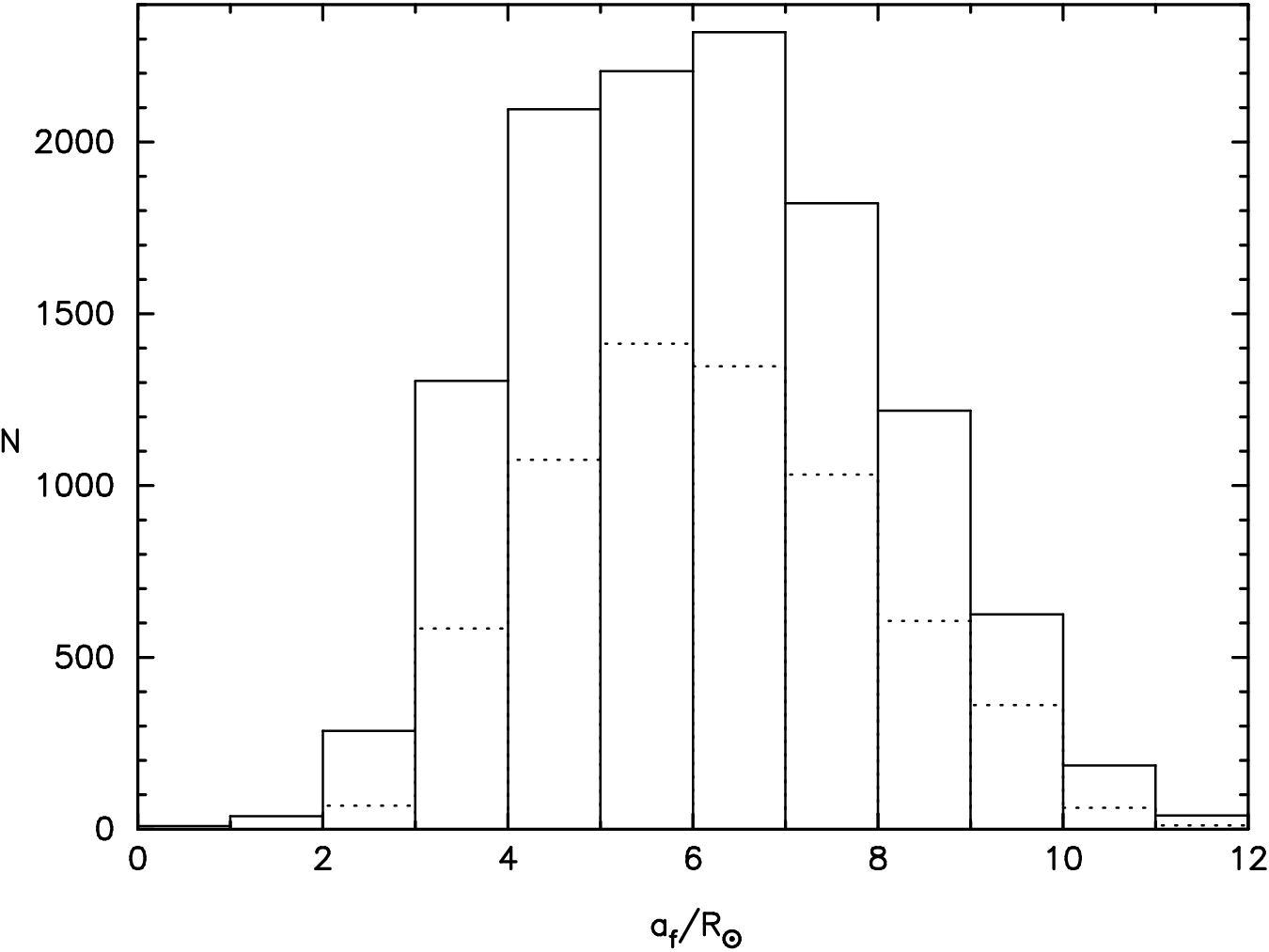}
   \caption[]{Model distributions over separation of components
 in the population of precursors of bh+ms binaries with $M_{\rm bh} \leq12$\,\ms,
$M_{20} \leq 1.2$\,\ms, $P_{\rm orb} < 1.2$ day after circularisation
of the orbits (upper panel) and in the ensemble of bh+ms binaries that produce
LMXB (lower panel). Solid line
-- \al=2, dotted line -- \al=0.5.}
      \label{fig:a}
}
\end{figure}

\begin{figure}
\parbox[t]{\columnwidth}
{ \includegraphics[angle=-90,width=0.98\columnwidth]{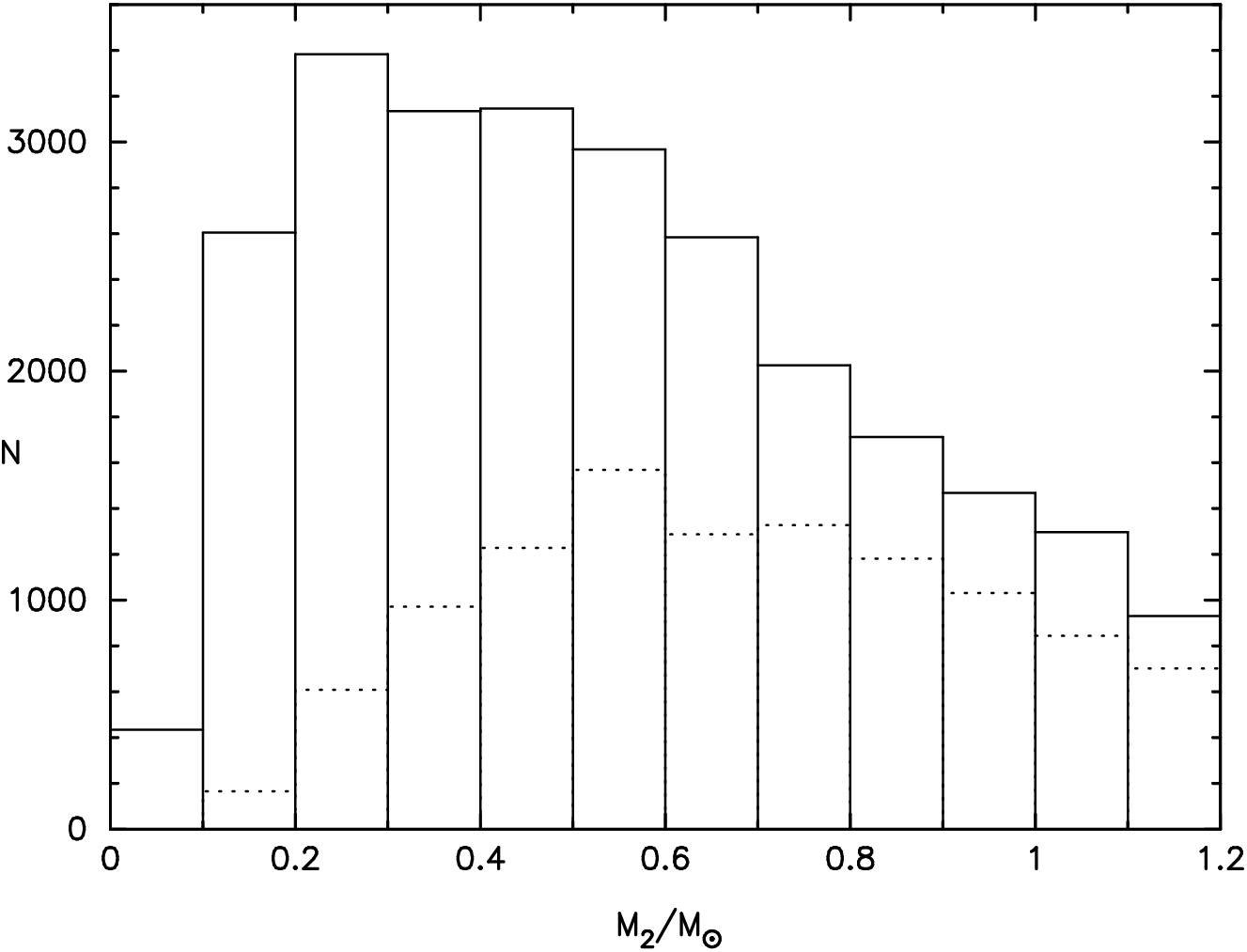}
    \includegraphics[angle=-90,width=0.98\columnwidth]{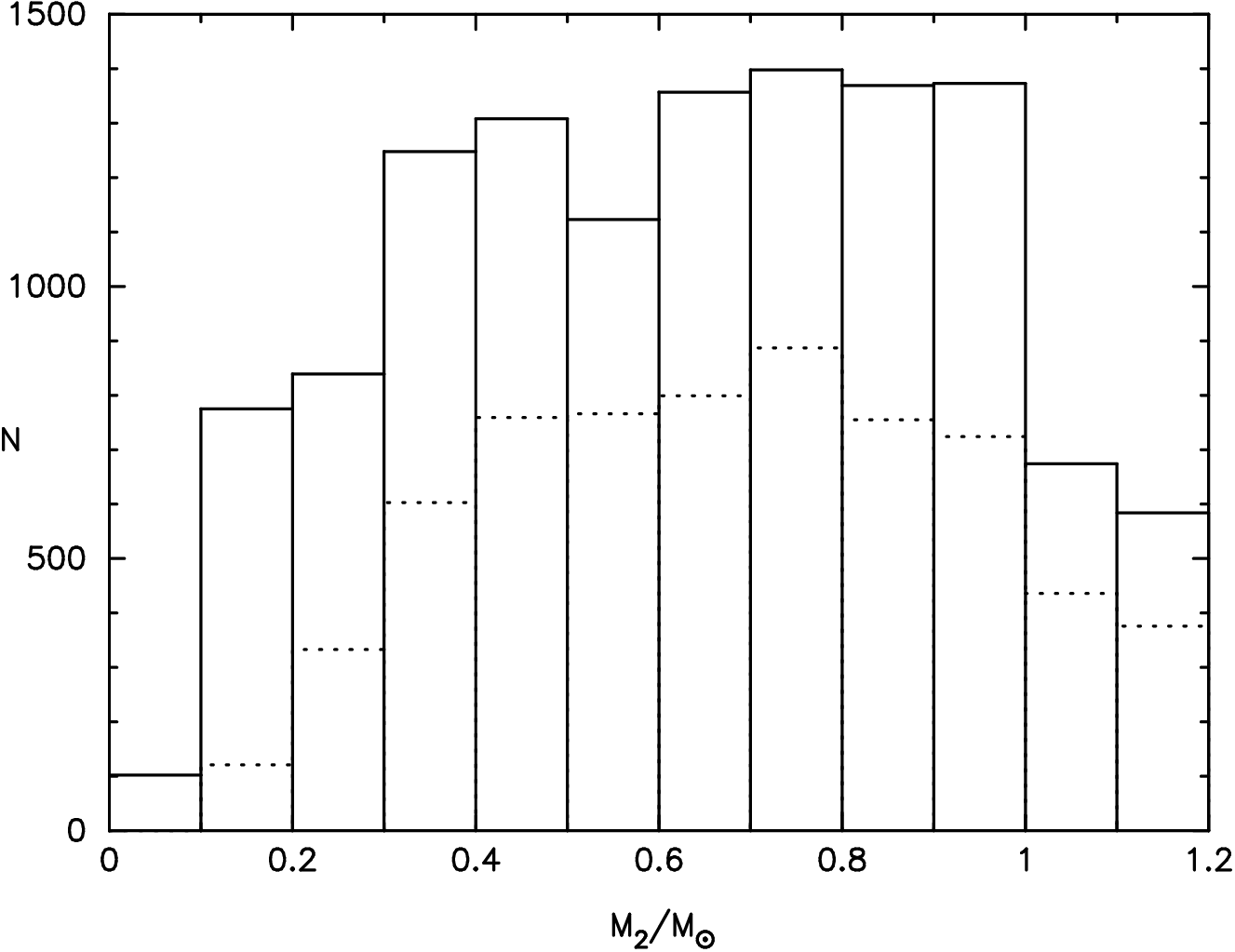}
   \caption[]{Model distributions of initial masses of secondary components in the
 population of precursors of bh+ms binaries with $M_{\rm bh} \leq12$\,\ms,
$M_{20} \leq 1.2$\,\ms, $P_{\rm orb} < 1.2$ day after circularisation
of the orbits (upper panel) and in the ensemble of bh+ms binaries that produce
LMXB (lower panel). Solid line
-- \al=2, dotted line -- \al=0.5.}
      \label{fig:m2}
}
\end{figure}

\begin{figure}[ht!]
    \includegraphics[angle=-90,width=\columnwidth]{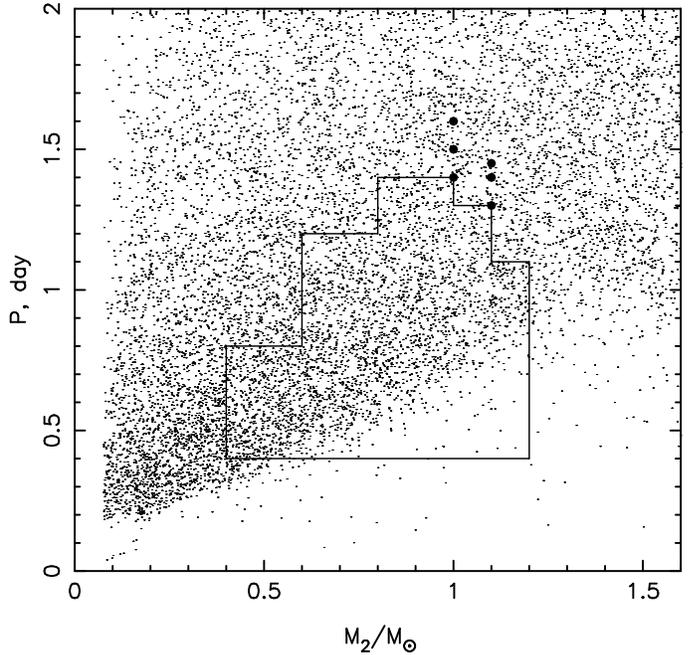}
   \caption[]{Galactic zero-age population of low-mass black hole binaries (dots).
The evolution of the systems that form LMBHB is determined by interpolation in the grid
of evolutionary tracks with the border outlined by the polygon or analytically, if
$M_{20} < 0.4$\,\ms\ and
 $P_0 <0.8$\,day.
Heavy dots mark initial parameters of the tracks listed in Table
 \ref{tab:tracks} and shown as
   examples in Figs. \ref{fig:pte} -- \ref{fig:pm2}}.
      \label{fig:grid}
\end{figure}

As discussed above, indirect estimates of the possible range of the
product of
common envelope equation parameters extend to $\al\aplt 2$.
We note that \cite{Taudew} have shown that, depending on the
definition of the core of the star and the treatment of the
role of internal thermodynamic energy, the binding energy
parameter $\lambda$ may vary by two orders of magnitude (from
0.02 to 3.50 for the same 20\,\ms\ star at the tip of red
giant branch) and therefore we consider $\al>1$ as an
acceptable value.
In Paper I we presented results of the modelling of
the population of LMBHBs
assuming a value of $\al=2$. Here, we
also discuss models with \al=0.1 and 0.5.

A run of \textsf{SEBA}-code with $\alpha_{\rm CE}\times\lambda =0.1$
produced about 3400 zero-age bh+ms binaries formed in a Hubble time.
However, all $M_2 \aplt 1.5\,\ms$ secondaries (i.e., stars subject
to magnetic braking) were paired with $\gta 14$\,\ms\ primaries,
exceeding the largest estimate of black hole mass in a known LMBHB
\citep[$9.7\pm0.6$ for A0620-500,][]{2007ApJ...663.1215F} -- and
often exceeding the largest dynamically evaluated mass of black hole
in binaries in the Galaxy
\citep[$15.65\pm$1.45\,\ms\,][]{2007Natur.449..872O}. \footnote{The
34 hr binary X-1 in the starburst galaxy IC 10 contains a $\gta
23\,\ms$ black hole \citep[][]{2008arXiv0802.2716S}.} Since this
model contradicts observations, we do not consider it further.

The model with \al=0.5 and AML via MSW implies the presence in the
Galaxy of about 1700 semidetached bh+ms binaries with $q \geq
0.02$, 220 of which are bright and stable. Based on arguments
presented in Paper I we conclude that, as in the \al=2 case, in
conflict with observations, up to several dozen persistent LMBHBs
then would be observed in the Galaxy above the \textit{RXTE All Sky
Monitor} sensitivity limit, thus implying the need for a reduced
MSW. The luminosities of these persistent LMBHBs are not high enough
to correspond to the bright steady X-ray sources observed in
elliptical galaxies \citep{2006MNRAS.371.1903I,2007arXiv0712.3052S}.

Thus, we consider results for
models in which AML via MSW does not act upon RLOF. Both
for \al=2 and 0.5, almost all semidetached bh+ms binaries
descend from systems with $\mbh\le12$\,\ms\ and $M_2 \leq
1.2$\,\ms. Table \ref{tab:numb} compares the number of
systems for two cases.\footnote{Since we present one random
realization of each model, all numbers given are subject to
Poisson noise.} The total numbers of systems differ\sout{s} by
a factor $\sim 2$ and apparently neither
 contradicts observation--based estimates of
several hundred to several thousand objects
\citep{csl97,romani1998}, though an \al=0.5 model may be
more attractive.

\begin{table}
\caption[]{Numbers of Galactic detached bh+ms systems with
$M_{\rm bh} \leq12$\,\ms, $M_{20} \leq 1.2$\,\ms, $P_{\rm orb}
< 1.2$ day formed in Hubble time, systems that reached RLOF,
and current number of semidetached LMBHB with $q \geq 0.02$.}
\begin{tabular}{lcc}
\hline\hline
  & \al=2 & \al=0.5 \\
Total number of systems & 25685 & 10910 \\
Systems that reached RLOF & 12150 & 6150 \\
Systems currently at $q\geq0.02$ & 5080 & 2980 \\
\hline
\end{tabular}
\label{tab:numb}
\end{table}

In the $M_{10} = (25-40)$\,\ms\ range to which most of the
precursors of black holes in bh+ms binaries belong, $M_{10} \gg
M_{20}$ and the second term in the brackets in Eq. (\ref{eq:ce}) is
$\gg 1$. Then, roughly, after the common envelope stage one obtains
the relation: $a_f/a_0 \propto \al$. Thus, the transformation law
for component separation may be written, approximately, as $f(a_f)
\Delta a_f = C\times h(a_0) \Delta a_0$, where $h(a_0)$ and $f(a_f)$
are, respectively, the distribution of systems over orbital
separations prior to and after the common envelope stage. By virtue
of this relation, functions $f$ and $h$ must be similar. This is
illustrated in Figs.~\ref{fig:a} and \ref{fig:m2}.

For lower \ace, initially wider progenitor systems are sampled,
as shown in the upper panel of Fig.~\ref{fig:a}.
Initially too ``close'' systems merge,
but they are replaced by initially wider systems. The rate of
AML given by Eq.~(\ref{eq:msw}) is a function of orbital
separation ($\dot{J} \propto a^{-0.5}$) and as a result
post-common-envelope separations of components in progenitors
of bh+ms systems that evolve into contact may descend only from
a very narrow range of $a$, given a limited time-span from
formation to the Hubble- or MS-evolution
times. In this range of $a$ the systems are distributed
similarly irrespective of \al\ (Fig.~\ref{fig:a}, lower panel).

Figure \ref{fig:m2} shows that, while for \al=2 initial
systems with low $M_2$ dominate, ``successful'' progenitors of
LMBHB have similar distributions over $M_2$ both for \al=0.5
and 2, since $\dot{J} \propto M_2R_2^4$. Masses of black holes
in the two \al\ cases also have similar distributions,
since for (25-40)\,\ms\ stars that are typical of most of the
progenitors of black holes in LMBHB, the masses of their
He-cores do not change significantly in the hydrogen-shell
burning stage that preceeds RLOF. As a combined effect of
similar distributions of detached bh+ms progenitors of LMBHB
over \mbh\, $M_2$ and $a$, scatter diagrams (e.g., $M_2 - P$)
for populations of LMBHB are similar both for \al =
0.5 and 2 and differ only in the number of systems per ``unit
area'' (by factor $\sim 2$), since initial systems were sampled
from different ranges in $a_0$ and our assumed initial
distribution over separations is $\propto 1/a_0$. For
consistency with Paper I we further consider the model for
\al=2.

Figure \ref{fig:grid} shows the distribution of zero-age ms+bh
binaries in the initial-mass-of-the-donor $M_{20}$ -- initial
 (post-circularisation of the orbit)
period $P_0$
plane.\footnote{Each dot in the plot
represents several systems that have similar $M_{10},
M_{20}, P_0$ but were born at different epochs in the history
of the Galaxy. Thus, some of the systems shown in the plot
 do not have time to evolve
into contact.} If a bh+ms system reached contact, its further
time-dependent behaviour was determined by interpolaton in the grid
of pre-computed evolutionary tracks. The borders of this grid are
outlined in Fig. \ref{fig:grid}. The initial systems with masses
larger than the rhs side border of the outlined range or periods
longer than the upper border of this region evolve to longer periods
upon RLOF and have unstable discs unless their mass is $\apgt
4$\,\ms. Evolution of the latter systems is illustrated by some
evolutionary tracks in the Figures \ref{fig:pte} and \ref{fig:pm2}.
Low-donor-mass systems, if their periods are not short enough, never
evolve to contact. There is, however, a contribution to the
population of LMBHBs from stars with $M_{20} < 0.4$\,\ms\ and
initial orbital period
 $P_0 <0.8$\,day and these systems were evolved analytically.
In binaries with $M_{20} \lta 0.6$\,\ms\ and
 $P_0 \gta 0.8$\,day the donors do not fill their Roche lobes in the Hubble time.
The initial masses of the progenitors of the donors are typically
$\lta 1$\,\ms\ and this means that most of the donors have to be
unevolved or slightly evolved at the instant of RLOF.

\section{Observational parameters of short-period LMBHBs}
\label{sec:obs}

\subsection{Effective temperatures}
\label{sec:teff}

\begin{table}[t!]
\caption[]{Parameters of tracks shown in the Figures (from left to right in each Figure. The
columns list the initial mass of the star, initial orbital period of the system, period at RLOF,
central hydrogen abundance at RLOF, the age of star at RLOF. In system 1 the
initial mass of the accretor is 12\,\ms, in systems 2 -- 8 the initial
mass of the accretor is 4\,\ms. }
\begin{center}
\begin{tabular}{rccccc}
\hline\hline
 No. & $M_0/\ms$ & $P_0,$ & $P_c,$ & $X_c$ & $T_c$\\
     & &day & day & & Gyr \\
\hline
 1 & 1.0 &0.4 & 0.316 & 0.696 & 0.07 \\
 2 & 1.0 &1.4 & 0.375 & 0.403 & 4.33 \\
 3 & 1.0 &1.5 & 0.395 & 0.324 & 5.30\\
 4 & 1.0 &1.6 & 0.420 & 0.231 & 6.33 \\
 5 & 1.1 &1.3 & 0.425 & 0.447 & 2.59 \\
 6 & 1.1 &1.4 & 0.446 & 0.369 & 3.22 \\
 7 & 1.1 &1.45 & 0.460 & 0.325 & 3.51 \\
 8 & 1.0 &1.9 & 1.772 & $8\cdot10^{-5}$ & 9.17 \\
\hline
\end{tabular}
\end{center}
\label{tab:tracks}
\end{table}

\begin{figure}
  \includegraphics[angle=-90,width=\columnwidth]{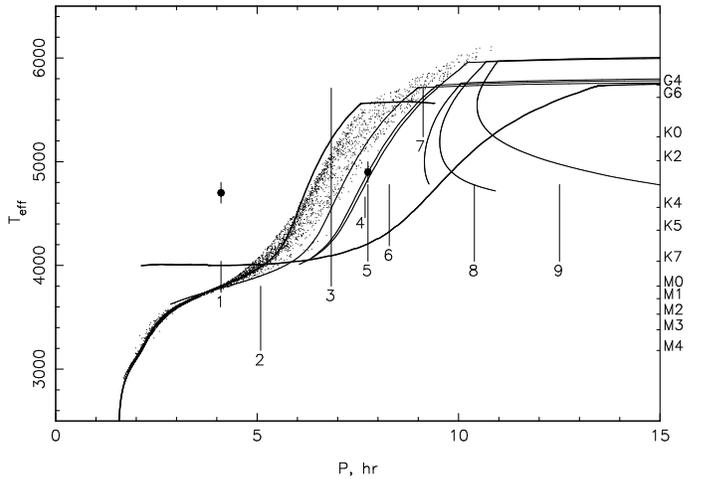}
   \caption[]{Model population vs. observational estimates of the ranges of effective
   temperatures of donors in SXTs (dots). Vertical lines mark the ranges of effective
   temperatures of
   donor-stars in observed SXTs corresponding to the ranges of the estimates of their
   spectral types (Table~\ref{tab:pte}). Systems are annotated according to their number
   in Table~\ref{tab:pte}. Large filled circles give $T_{\rm eff}$ of donors derived
   from the fits to synthetic spectra. Heavy solid lines to the left and right show ``limiting'' tracks
   for a (1+12)\,\ms, $P_0=0.4$ day system in which MB does not operate after RLOF and
   a (1+4)\,\ms, $P_0=1.9$ day system with MB operating after RLOF (see \S \ref{sec:teff}
   for discussion). Thin solid lines are evolutionary tracks for
   $M_0=1.0$ and $1.1$\,\ms\ donors with 4\,\ms\ accretors listed in Table \ref{tab:tracks}.
   The {\sl Sp} $-T_{\rm eff}$ relation used in the paper is shown at the right
   border of the coordinate box.
              }
     \label{fig:pte}
\end{figure}

As noted by \citet{2006MNRAS.366.1415J}, any formation scenario for
LMBHBs has to explain the spectral types of black-hole low-mass
companions. The determination of spectral types is a
challenging task since the emission of the cool star is
contaminated by the radiation from the accretion disc and the
hot spot where the accretion stream hits the disc's edge
\citep{2006csxs.book..215C}. Moreover, contamination by the
disc may vary with time if the system did not reach quiescence.
The published methods of spectral type determination vary in
sophistication from naked eye estimates to using $\chi^2$
statistics after subtracting template spectra from
Doppler-corrected averaged spectra \citep[see, e.g.,
][]{1996PASP..108..762H};
sometimes spectral types are inferred from colours, absolute
magnitudes, SED and in many cases are subjective. For this
reason, estimates of the donor's spectral type for the same
system may differ by several subtypes (see Table \ref{tab:pte}
where we summarised the published spectral types for LMBHBs).

The effective temperature of the donor is available in the
literature for XTE~J1118+480 (KV UMa) and A0620-00 (V616~Mon)
only. They were derived as a by-product of abundance
determinations that used synthetic spectra and
$\chi^2$-minimisation techniques\footnote{
\citet{2007ApJ...663.1215F} cast doubt upon the temperature
determination for A0620-00 by \citet{2004ApJ...609..988G},
claiming that the latter authors used an insufficient set of
spectral lines in their study and overestimated \te.}. For
other systems we were forced to apply the {\sl Sp} $-T_{\rm
eff}$ relation for zero-age
main-sequence stars. We used the relation given
by Tokunaga in \citet{2000asqu.book.....C}. This relation is
accepted by the authors of the catalogue of Hipparcos
spectroscopic standards. For A0-M6 spectra the approximate
uncertainty of this scale (one standard deviation) is
$\pm100$\,K.

A caveat has to be entered concerning observers assigning
``nonexistent'' spectral subtypes to their objects. The modern MK
classification system is devised in such a way that subsequent
subtypes represent approximately equal differences in the spectra
and some original decimal subdivisions were dropped. For instance,
some subtypes between K5 and M0 are absent: ``K7 is considered as
half a subtype later than K5 and earlier than M0''
\citep{1985IAUS..111..121K}. For spectral types absent
in the spectral classification (e.g., K3V) we
applied a linear interpolation in $\log T_{\rm eff}$.

In Fig. \ref{fig:pte} we plot the distribution of our model
population in the $P_{\rm orb}-T_{\rm eff}$\ plane and compare
it with the $T_{\rm eff}$ of particular systems. In the absence
of objective criteria for the discrimination of reliable vs.
non-reliable spectral type determinations, we plot for each
system the complete range of the effective temperature for the
range of spectral types assigned to it by different observers.
Within the uncertainties of the spectral type determinations and
conversion
{\sl Sp} $-\te$
the model satisfactorily reproduces
\te\ of the donors in the LMBHBs with $P_{\rm orb} \lta9$\ hr.

\citet{2003MNRAS.340.1214P}
noticed that for $\te \lta 4500$\,K stellar models with grey
atmosphere boundary conditions tend to overestimate the
effective temperatures of stars, compared to models with more
realistic non-grey atmospheres. Using
unevolved models of low-mass stars, Podsiadlowski et al.
estimated that the correction in temperature may amount to
about 350\,K. However,
(i) they used
a different stellar structure code, (ii) the correction may be
different if the comparison is made for non-grey atmosphere
models of out-of-thermal-equilibrium mass-losing stars.
However, models of the latter kind have not yet been computed
and for the moment the difference between our models and the
non-grey atmosphere models cannot be estimated.

Some systems -- GRS~1009-45, XTE~1650-500, A0620-00, and GS
2000+25 -- are apparently located below the ``populated'' area.
However, we restricted the initial periods of
bh+ms binaries and masses of donors with the limits shown by
the solid line in Fig.~\ref{fig:grid} so that interpolation
between pre-computed tracks that upon RLOF evolve continuously
to shorter periods and tracks that immediately go to longer periods
(or change the direction of evolution in $P_{\rm
orb}$) is avoided. As can be seen from the tracks plotted in
Fig. \ref{fig:pte} and from Table \ref{tab:tracks} (tracks 2 -
7), for a system with a given $M_{10}$ and $M_{20}$ the
direction of evolution changes quite abruptly over a narrow
range of initial $\Delta \per \lta 0.1$\ day. For a given
combination of $M_{10}$ and $M_{20}$ a ``gap ''
between tracks evolving in different directions forms. But
since there is a continuity in the initial parameters of the
systems, the ``gap'' in reality is filled. This is
clearly shown by the tracks plotted in Fig. \ref{fig:pte}.
Initial parameters of these additional tracks
belong to a well populated area in the $M_{20} - P_0$ diagram (Fig.
\ref{fig:grid}). We did not pursue the goal of finding the precise
parameters of the initial system(s) that may fit the parameters of a
particular observed system and the precise borders of the progenitor
space, since the accuracy to which the parameters of SXT are known
and the uncertainty in the efficiency of magnetic braking do not
justify this time-consuming and computationally expensive task.

 Nevertheless it is clear, at least
qualitatively, that the origin of short-period LMBHBs may
be explained within the paradigm of the strongly reduced magnetic
braking in systems with donors overflowing Roche lobes.

In Paper I we reduced the AML by magnetic braking (MB) to 0. This
might be excessive and in reality some amount of MB can be still
operating. (But we found that reducing the MB by a factor of 2
still leaves some $\sim 100$ bright steady sources.) We plot in Fig.
\ref{fig:pte} two ``limiting '' tracks: for (1+12)\,\ms,
$\per_0=0.4$ day in which the donor is almost unevolved at RLOF and
MB is absent after RLOF and for (1+4)\,\ms, $\per_0=1.9$ day in
which the donor has $X_c \simeq 10^{-4}$ at RLOF and MB continues to
operate (tracks 1 and 8 in Tab. \ref{tab:tracks}). Crudely, model
populations with MB and without MB have to be located between these
two limiting curves. Of course there will be a contribution from
lower and higher mass systems, as we plotted only the 1\ms\ tracks
for simplicity.
 Therefore adding some MB to our model will shift the
population to the right, reaching a better agreement with
observations while still not producing stable luminous sources.
From Fig. 6 in Paper I one can see that such an addition of MB
mainly will influence the long-period systems.

From Table \ref{tab:tracks} and Fig. \ref{fig:pte} one can
see that stars turn to longer periods if $X_c \lta 0.35$ at
RLOF. For instance a 1.1\ms\ donor with $\per_0=1.4$\
(second to the right line)
spends almost 10 Gyr in the RLOF-state, out of
which during about 5 Gyr it evolves to longer periods. This
provides the possibility of explaining SXTs with periods $>$
9-10 hr. However, it is then necessary to compute a new grid of
tracks for stars that evolve to longer periods and a very dense
grid of tracks to cover the space between tracks evolving to
shorter and longer \per. This will be the topic of a dedicated paper.

\subsection{Mass transfer rates}
\label{sec:dmdt}

\begin{figure}
    \includegraphics[angle=-90,width=\columnwidth]{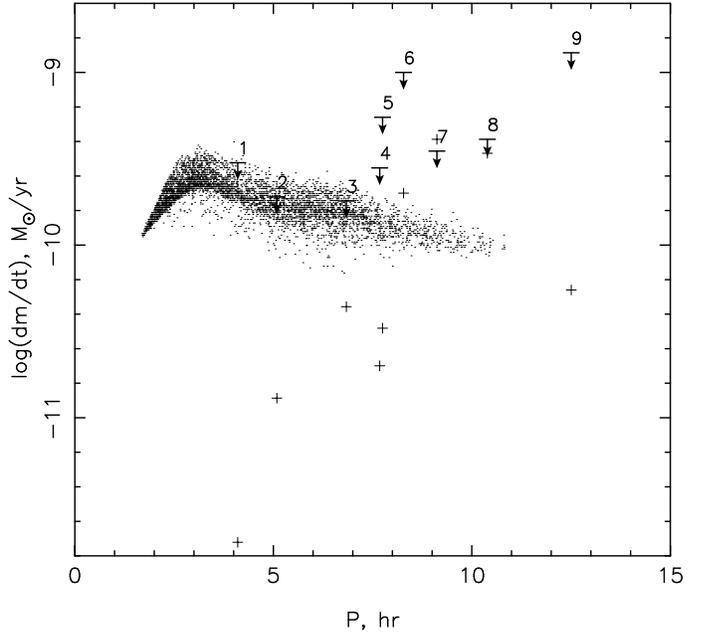}
   \caption[]{ Mass transfer rates in model LMBHB as a function of their orbital periods (dots).
Arrows mark
upper limits to the estimates of mass-transfer rates in observed
SXTs as given by Eq. (\ref{eq:dmdtmax}). Crosses are estimates of
mass-transfer rates in SXTs based on recurrence times.
 }
       \label{fig:pdmdt}
\end{figure}

There is no secure method of mass transfer rate
determination for transient LMXBHs. One can obtain an estimate
of this parameter by dividing the mass accreted during outburst
by the recurrence time. However, this approach has several
weaknesses. First, recurrence times are known only for a
few systems. Second, it is not sure that the rate calculated in
this way
represents the secular value (this is the general drawback of
mass-transfer rate estimates) and third, it assumes that during
the ``refill", accretion onto black hole does not occur.
This last assumption is
put in doubt both by observations \citep[see e.g.][ and
references therein]{2007A&ARv..15....1D} and models
\citep[see][ and references therein]{2008arXiv0801.0490L} which
suggest that quiescent discs are truncated and therefore leaky.
In such a case one can estimate the upper limit to the
mass transfer rate which cannot be larger than the
critical-for-stability accretion rate at the truncation radius
(see Paper I for details). The actual mass transfer rate should
be somewhere between the values estimated by the two methods.

Figure \ref{fig:pdmdt} compares the model mass transfer rates
with observational estimates of \md. We present two estimates
of the latter. We show \md\ estimates from recurrence times and
mass accreted during the outburst and estimates of the upper
limit of the accretion rate at the truncated
disc inner edge. In the latter case, we get:
\begin{eqnarray}
\label{eq:dmdtmax}
\lefteqn{\md_{\rm max} \lta} \nonumber\\
 & 2.5\cdot 10^{-7}
\left[\left(1+q\right)^{1/3}\left(0.5-0.227 \log q\right)\right]^{10.32} P_d^{1.72}f_t^{2.58}\myr.
\end{eqnarray}
In Eq. (\ref{eq:dmdtmax}) $P_d$ is the orbital period in days, and
$f_t \lta 0.48$ is the fractional disc truncation radius.
 We
used the revised version of the critical accretion rate
\citep{2008arXiv0802.3848L}:
\begin{equation}
\dot{M}_{\rm
crit}^{-}=2.64\times10^{15}~\alpha_{0.1}^{0.01}~R_{10}^{
2.58}~M_1^{-0.85}\,\rm g\,s^{-1}.
\end{equation}

The estimates of mass-transfer rates for leaky discs differ
slightly from the ones given in Table 3 of Paper I, since in
the present study we used in the equation for $\md_{\rm max}$
the lower limit of $q$ as given in Table \ref{tab:pte} instead
of assuming a similar $q=0.1$ for all systems. For V406 Vul
we estimated $q$ by using the value of the mass function and
the mass of the secondary as obtained from its spectral type.

The estimates of $\md_{\rm max}$ for XTE J1118+480, GRO J0422+32,
and GRS 1009-45 strongly suggest that the AML in short-period LMBHB might
be defined by GWR only.

H1705-25 (system 9) may, as noted in the previous
subsection, belong to the population in which donors fill Roche
lobes when their central hydrogen abundance $X_c$ is reduced by
$\gta$50\%.

\subsection{Masses of secondaries}
\label{sec:m2}

\begin{figure}
    \includegraphics[angle=-90,width=\columnwidth]{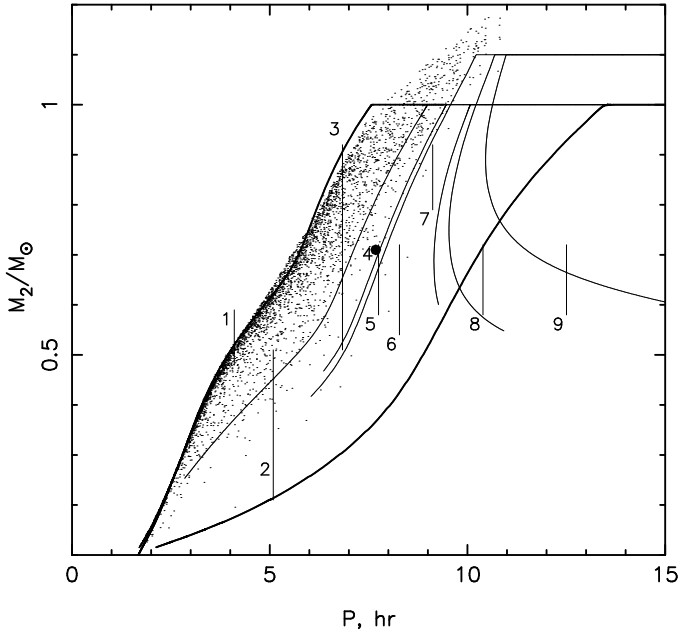}
   \caption[]{Masses of donor-stars in modelled population. Vertical
lines show the ranges of $M_2$
corresponding to the uncertainty in the determination of spectral types (Table \ref{tab:pte}).
Heavy solid lines
   show ``limiting'' tracks as in Fig. \ref{fig:pte}.
Thin solid lines are evolutionary tracks for 1.0 and 1.1\,\ms\ donors with 4\,\ms\ accretors like in Fig. \ref{fig:pte}.
For XTE~J1650-500 (at \per=7.88 hr) there is only one determination of spectrum -- K4V, but
the same authors \citep{2004ApJ...616..376O} mention that the next best fits are G5V and K2III; for this reason we
show the lower limit for $M_2$ in this system.}
     \label{fig:pm2}
\end{figure}

Figure \ref{fig:pm2} compares the ranges of the masses of
donors in observed SXTs corresponding to the ranges of the
estimates of their spectral types with the masses of donors in the
model. The spectrum-mass scale is adopted after Schmidt-Kaler, as given
in \citet{2000asqu.book.....C}.

The situation with $M_2$ is similar to that with \te: our model
population well covers the masses of the four shortest period
systems, but to explain longer period ones we need to apply
tracks for more evolved systems that we did not include in
our grid of tracks. Adding some AML due to MB also would
improve the agreement with observations. Figure~\ref{fig:pm2}, like
Fig. \ref{fig:pte}, suggests that the origin of LMBHBs with
orbital periods of 10 -- 12 hr may be associated with systems
in which RLOF occurred when $X_c \lta 0.35$.

\section{Discussion and conclusion}
\label{sec:concl}

We have shown above that assuming the product of the
stellar-envelope binding-energy parameter and the
common-envelope expulsion efficiency parameter \al=2,
it is possible to reproduce, (within the uncertainty of
observations) the number of LMBHBs in the Galaxy, the effective
temperatures and masses of the donors in these systems (as
inferred from the spectra of the latter) and their
mass-transfer rates. This result is maintained for
\al=0.5 but further reduction of
\al\ to 0.1 results in models whose parameters are not
compatible with the currently available data on observed
systems. Also we reiterate that (as found in Paper I) a
substantial reduction of the strength of magnetic braking as
compared to the ``standard'' \citep{vz81} makes all calculated
systems transient, in agreement with observations.

The common envelope phase remains the most enigmatic phase of
binary star evolution. As long as the processes of interaction
of the companion star with the envelope it is penetrating is not
understood, the use of simple conservation-law based
equations will remain the necessary, albeit approximate,
approach. However, the evolution
of massive stars strongly depends also on stellar winds. The
mass of the black-hole progenitor and the mass of its envelope, and
the radius of the star that define the outcome of the common
envelope stage, are interrelated via mass loss in the
pre-common-envelope stage, which is not well constrained. The
survival of a binary in a supernova explosion depends on the
mass-loss in the Wolf-Rayet star phase, the possible kick
imparted to the nascent black hole and the fraction of the mass of
the exploding star that forms the black hole. None of these parameters
are well constrained.

The situation concerning mass-loss by massive stars
is controversial. It became a recognized fact that stellar
winds of both O-stars
 and WR-stars
 are clumped
\citep[e.g.,][]{1988ApJ...335..914O,2007arXiv0710.3430S,1988ApJ...334.1038M,hamann_koest_wr98}
and
empirical estimates of mass-loss rates that depend quadratically on the density
have to be revised downward by a
factor of several.
  In particular, downward revision of empirical \md\ values would bring
them into agreement with modern theoretical \citep{vkl01}
rates for OB-stars
\citep{2007A&A...473..603M,2007AIPC..948..389V,2007arXiv0708.2066V}.
It is also claimed that the widely accepted \citet{nl00} rates for
WR-stars, which already are clumping-corrected, have to be
revised further downward \citep{hamann_wn06}. A decrease
of \md\ would mean more massive stellar envelopes and, generally,
more energy would be needed for the
ejection of common envelopes. This may be interpreted as a need for higher \ace.

On the other hand, based on results of extensive analysis
of the ratios of blue to red supergiants, of Wolf-Rayet stars
to O supergiants, of red supergiants to Wolf-Rayet stars and of
the relative number of Wolf-Rayet subtypes, WC to WN stars,
\citet{2008MNRAS.384.1109E} suggested that the total
amount of mass lost by stars has to be increased. In the latter
study both single and binary evolutionary models were
considered and the mass-loss rates of \citet{vkl01} and
\citet{nl00} were used.

In such
a controversial
 situation we
may only claim, based on our results, that for the
parameters of stellar evolution
implemented in our evolutionary and population synthesis codes,
which are consistent with state-of-the art stellar evolution theory,
an agreement between the properties of the observed population
of short-period LMBHBs and the model suggests a high efficiency
of expulsion of common envelopes.

Above, we have shown that the number and
properties of observed short-period LMBHB may be explained by
the model that assumes a strongly reduced strength of magnetic
braking. But we recall the existence of the
alternative model of \citet{mnl99} which suggests that most
LMBHB might have truncated discs that are
secularly in a cold and stable equilibrium with transiency due
to random variability in the properties of
discs and/or mass transfer rates. Within this
model most of the systems we modeled and classified as transients
may appear as faint and stable. At the
moment there are no theoretical arguments against the Menou et
al. model and observations that may serve as
selection criterion between the two models
still have not been defined. Thus, both
models remain possible.

\begin{acknowledgements}
We thank P.P. Eggleton for providing the copy of his
evolutionary code. We acknowledge Gijs Nelemans for providing
models of zero-age populations of ``black-hole + main-sequence
star'' binaries and numerous discussions, Guillaume Dubus
for the analysis of stability of accretion disks
and Lev Titarchuk and Nikolai Shaposhnikov
for useful comments. LRY acknowledges the warm hospitality and
support from the Institut d'Astrophysique de Paris, Universit\'e
Pierre et Marie Curie where a part of this study was carried
out. JPL and LRY are grateful to Phil Charles and Hans Ritter
for helpful discussions at the Trzebieszowice Castle and to
Saul Rappaport and Philip Podsiadlowski for their illuminating
critique of Paper I. We thank the anonymous referee for
a thorough reading of the manuscript and useful comments.
This work was supported by the Centre National d'Etudes
Spatiales (CNES) and by a grant from the CNRS GDR PCHE. LRY is
supported by RFBR grant 07-02-00454 and Russian Academy of
Sciences Basic Research Program ``Origin and Evolution of Stars
and Galaxies''.
\end{acknowledgements}


\end{document}